\documentclass{article}

\usepackage{fancyhdr}
\usepackage{extramarks}
\usepackage{graphics}
\usepackage{listings}
\usepackage{xcolor}
\usepackage{subfig}
\usepackage{soul}
\usepackage{amssymb}
\usepackage{float}
\usepackage[final]{pdfpages}
\usepackage[fleqn]{amsmath}
\usepackage{booktabs,siunitx}
\usepackage{ragged2e}

\definecolor{dkgreen}{rgb}{0,0.6,0}
\definecolor{gray}{rgb}{0.5,0.5,0.5}
\definecolor{mauve}{rgb}{0.58,0,0.82}

\title{
Traffic prediction at signalised intersections \\ using Integrated Nested Laplace Approximation
}

\author{\textbf{Dale Townsend and Charne Nel}}

\begin{document}

\maketitle
\newpage

\pagenumbering{roman}
\tableofcontents
\newpage
\pagenumbering{arabic}

\section{Abstract}

A Bayesian approach to predicting traffic flows at signalised intersections is considered using the the INLA framework. INLA is a deterministic, computationally efficient alternative to MCMC for estimating a posterior distribution. It is designed for latent Gaussian models where the parameters follow a joint Gaussian distribution. An assumption which naturally evolves from an LGM is that of a Gaussian Markov Random Field (GMRF). It can be shown that a traffic prediction model based in both space and time satisfies this assumption, and as such the INLA algorithm provides accurate prediction when space, time, and other relevant covariants are included in the model.

\section{Introduction}

Hamilton City Council currently operates 104 signalised intersections in the Waikato region. Traffic counts are recorded at each of these using a device called a Detector in each lane of the oncoming roadways. Decisions for changes to infrastructure are made taking traffic flow into account, and as such it is important to accurately predict future traffic flow. In addition the detectors often fail, resulting in missing traffic counts at various times. This missing data is explored as part of the model fitting and the fitted values from the model output is considered as a form of imputation. \newline

The goal of the model is to find the variables with a large influence on the traffic count, and use them to fit a prediction model in INLA. These variables are explored to gain an understanding of their use as covariates in the model. Prediction error for each model is obtained to distinguish between the prediction accuracies of the respective models.

\section{INLA}
The need to efficiently analyse large datasets of spatial and temporal models has increasing importance with the recent advances in big data collection. INLA is a computational, deterministic algorithm designed for efficient inference of models with a high spatial and temporal resolution. While MCMC models have traditionally been used for this class of models, there is a high computational load when dealing with large datasets. Instead of the simulation based algorithm used by MCMC, INLA is an analytic approximation based on the Laplace method.

\subsection{Latent Gaussian Models}
A latent field comprises of unobserved variables which is used to infer on a set of parameters $\theta$ $=$ \{$\beta_{0}$, $\beta$, $f$\}. In order to define a latent Gaussian field,  the observed data needs to be specified: y $=$ (y$_{1}$, $\cdots$, y$_{n}$). Each y$_{i}$ is characterised by some parameter $\phi_{i}$, which can be defined by an additive structured predictor $\eta_{i}$ through the link function. The linear predictor $\eta_{i}$  is given as:
\begin{equation}
\eta_{i} = \beta_{0} + \sum_{m=1}^{m}\beta_{m}x_{mi} + \sum_{l=1}^{L}f_{l}(z_{li})
\end{equation}
$\beta_{0}$ is a scalar that acts as the intercept in the model, $\beta$ relates to the coefficients that quantify the effects of the covariates $x$ in the model, and $f$  is a collection of the functions that are defined as a result of the set of covariates $z$. As mentioned, the latent field components are collected in a set of parameters $\theta$, defined as $\theta$ $=$ ($\beta_{0}$, $\beta$, $f$) and the hyperparameters for the model is $\Psi$ $=$ \{$\Psi_{1}$, $\cdots$, $\Psi_{K}$\}. Two properties that are prevalent in most latent fields are the assumption of conditional independence, and that the number of hyperparameters in the model is relatively small. The data that are being explored is not necessarily normally distributed, but the latent field is. The likelihood of the $n$ observations in the data, assuming they are from the exponential family of distributions and are conditionally independent, is given by the likelihood

\begin{align}
p(y|\theta,\Psi) = \prod_{i=1}^{n}p(y_{i}|\theta_{i},\Psi),
\end{align}

where each observation $y_{i}$ is connected to only one element $\theta_{i}$ in the latent field of parameters $\theta$. [1]

\subsection{Gaussian Markov Random Fields}
The conditional independence property allows for the creation of a sparse precision matrix $\mathcal{Q}$($\Psi$). This feature within the latent field, gives rise to the Guassian Markov random field (GMRF). There is a significant computational benefit when GMRFs are used in models due to the sparse nature of the precision matrix. GMRFs also have the Markov property, which is needed for models that rely on MCMC sampling.
When areal data is used for a given area $i$, its neighbors $\mathcal{N}$($i$) are strictly the areas which it shares its borders with. The Markovian property in the context of a spatio-temporal model means that an observation $y_{i}$  for the $i$th area, is independent of all other observations, given the set of neighbors $i$ has. [1]
\begin{align}
y_{i} \perp y_{j}|Y_{\_ij}, i \neq j
\end{align}

\subsection{Laplace Approximation}
Laplace approximation is an alternative to simulation-based Monte Carlo integration, where the aim is to approximate the posterior. We are interested in computing the integral
\begin{align}
\int f(x)dx = \int exp(log f(x))dx
\end{align}
The Laplace approximation will find a Gaussian approximation to the conditional distribution of a set of continuous variables. The maximum of an integral is found, and then Taylor series approximation is applied to the logarithm of the function to calculate the Laplace approximation. The integral evaluated in the interval $\alpha, \beta$ is approximated by
\begin{align}
\int_{\alpha}^{\beta} f(x)dx \approx f(x^{*})\sqrt{2\pi \sigma^{2^{*}}}(\Phi{\beta})-\Phi(\alpha))
\end{align}
where $\Phi(.)$ represents the cumulative density function of the Normal$(x^{*}, \sigma^{2^{*}})$ distribution. In the case of a Gamma distribution, the mode $x^{*}$ is obtained by solving $\frac{\partial log f(x)}{\partial x} = 0$. The variance $\sigma^{2^{*}}$ is obtained by evaluating $-1/\frac{\partial^{2} log f(x)}{\partial x^{2}}$ at the mode $x^{*}$  [1]. The Laplace approximation is then given by 
\begin{align}
\text {Gamma}(a,b) \approx \text{Normal}(x^{*} = \frac{a-1}{b}, \sigma^{2^{*}} = \frac{a-1}{b^{2}})
\end{align}

\section{Data}

\subsection{SCATS Detectors}

Hamilton City Council controls all of the traffic lights in the city boundaries, including those on State Highways. The system used to manage them is called SCATS. SCATS is installed in 154 cities across 25 countries to manage the timing of signal phases at intersections. Devices called detectors are installed at the end of the each lane at a signalised intersection. They are inductive coils that send a message to the detector controller when a car passes over them. SCATS dynamically manages the timing of the signals in accordance with the number of vehicles passing over the detectors. For example at intersections with a primary and secondary road, during non peak hours the signals at the secondary road will stay red until a vehicle is waiting at it. \\

One form of data output from SCATS is the count of vehicles passing over each detector over a given time span. We will be using these counts to predict the number of vehicles at each detector of an intersection at a given time. The counts in a SCATS traffic count export represent the total number of vehicles that have passed completely across the detector. It will not count vehicles twice, unless a vehicle happens to reverse back past the detector and over it again. The detectors are placed just before the end of the lane so that turning vehicles can be distinguished from through vehicles. Although data is available for individual detectors, we will be using the sum of all detectors to summarise the counts into the total number of vehicles passing through the intersection in a given time span.

\begin{figure}[H]%
    \centering
    \subfloat[Detector placement at a signalised intersection (RAMM)]{{\includegraphics[width=8cm]{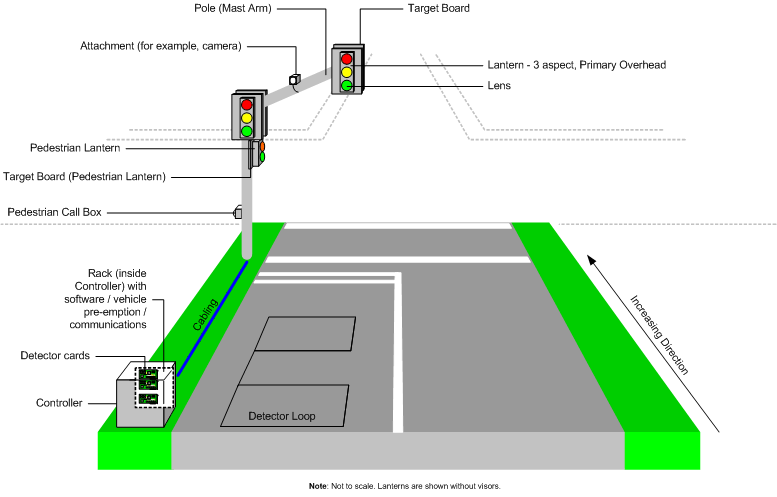} }}%
    \qquad
    \subfloat[Detector placement at Moonlight-Borman]{{\includegraphics[width=7cm]{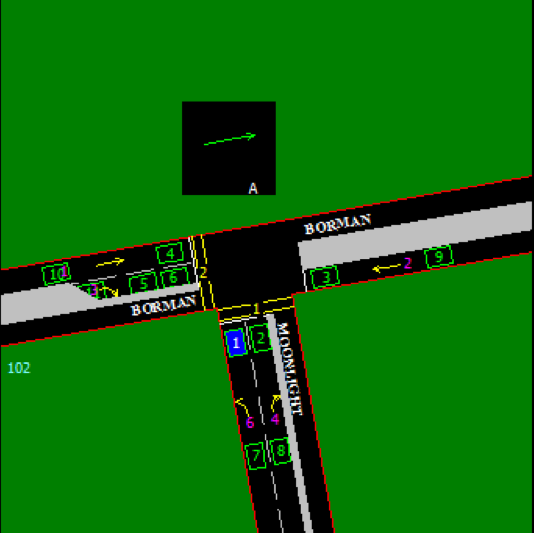} }}%
\end{figure}

\subsection{SCATS Data Cleaning}
The raw exports from SCATS require a large amount of cleaning before they can be used in the INLA model. As the data for each intersection includes the counts for all detectors, the sum of the detector counts is not necessarily the total number of vehicles passing through the intersection. To achieve this each site was examined for the detectors to keep. The above SCATS intersection diagram (right) shows the detector placement at an intersection in northern Hamilton. Detectors 7 and 1 generally count the same number of vehicles, as do 2 and 8, 3 and 9, 5 and 6, and 4 and 10. The detectors further back from the intersection are known as 'advance detectors'. For our purposes these are not needed. The detectors closest to the stop line count all traffic passing through the intersection, and as such all other detector counts were removed from the data. \\

A sum variable was added for each site. This is the sum of the counts at all remaining detectors. In most cases this represents the total vehicles passing through the intersection within each 30 minute interval. A site variable was also added for distinguishing between sites when the data was merged into one data frame. In addition sites outside the city boundaries were removed - Sites 50, 54, 76, 90, 91, 94, and 104. Site 52 is not a valid site and was also removed. The remainder is a total of 96 sites in the city boundary. 61 of these are at signalised intersections and the remaining 35 are pedestrian crossings. \\

The clean data for all sites was joined row-wise, such that the variables of the data are in columns and each observation is a row. A large amount of missing data was found. The detectors at times fail for various reasons, and when this occurs the count is recorded as 'BAD'. HCC currently use an Excel macro to impute some of this missing data, which copies over the data from the previous or following day. In the interest of data integrity we did not impute or remove any missing data. A date range with a large amount of missing data was compared to a range with substantially less, and the two models compared to understand the effect of the missing data on prediction accuracy. The results of this are explained further in Section *. \textit{Plot of NA per week}.

\begin{table}[H]
\centering
\caption{Example of the cleaned SCATS data}
\begin{tabular}{lllll}
\hline
\multicolumn{1}{|l|}{\textbf{Date}} & \multicolumn{1}{l|}{\textbf{Time}} & \multicolumn{1}{l|}{\textbf{Site}} & \multicolumn{1}{l|}{\textbf{Sum}} & \multicolumn{1}{l|}{\textbf{ID}} \\ \hline
16/10/17                   & 00:00-00:30               & 1                         & 90                       & 1                       \\ \hline
16/10/17                   & 00:30-01:00               & 1                         & 67                       & 1                      \\ \hline
\end{tabular}
\end{table}

\subsection{Data Exploration}

Prior to modelling an exploration was first carried out on the cleaned dataset to understand how the traffic count varies over time and space.

\begin{lstlisting}
> summary(data)
       X               Date                Time                Site             Sum        
 Min.   :     1   Min.   :2017-10-16   Length:855744      Min.   :  1.00   Min.   :   1.0  
 1st Qu.:213937   1st Qu.:2017-12-01   Class :character   1st Qu.: 24.75   1st Qu.:  78.0  
 Median :427872   Median :2018-01-16   Mode  :character   Median : 48.50   Median : 283.0  
 Mean   :428337   Mean   :2018-01-16                      Mean   : 50.51   Mean   : 428.7  
 3rd Qu.:641808   3rd Qu.:2018-03-04                      3rd Qu.: 75.50   3rd Qu.: 651.0  
 Max.   :864658   Max.   :2018-04-19                      Max.   :103.00   Max.   :2674.0  
                                                                           NA's   :78241   
\end{lstlisting}

Here 'Sum' represents the total traffic through a given intersection at a given time. There are a total of 864,658 observations. Of these, 74,241 of the Sum values are missing. The max Sum value is unusual in comparison to the previous quartile values. This could indicate a right skewed distribution.

\begin{figure}[H]
\centering
    \includegraphics[width=12cm]{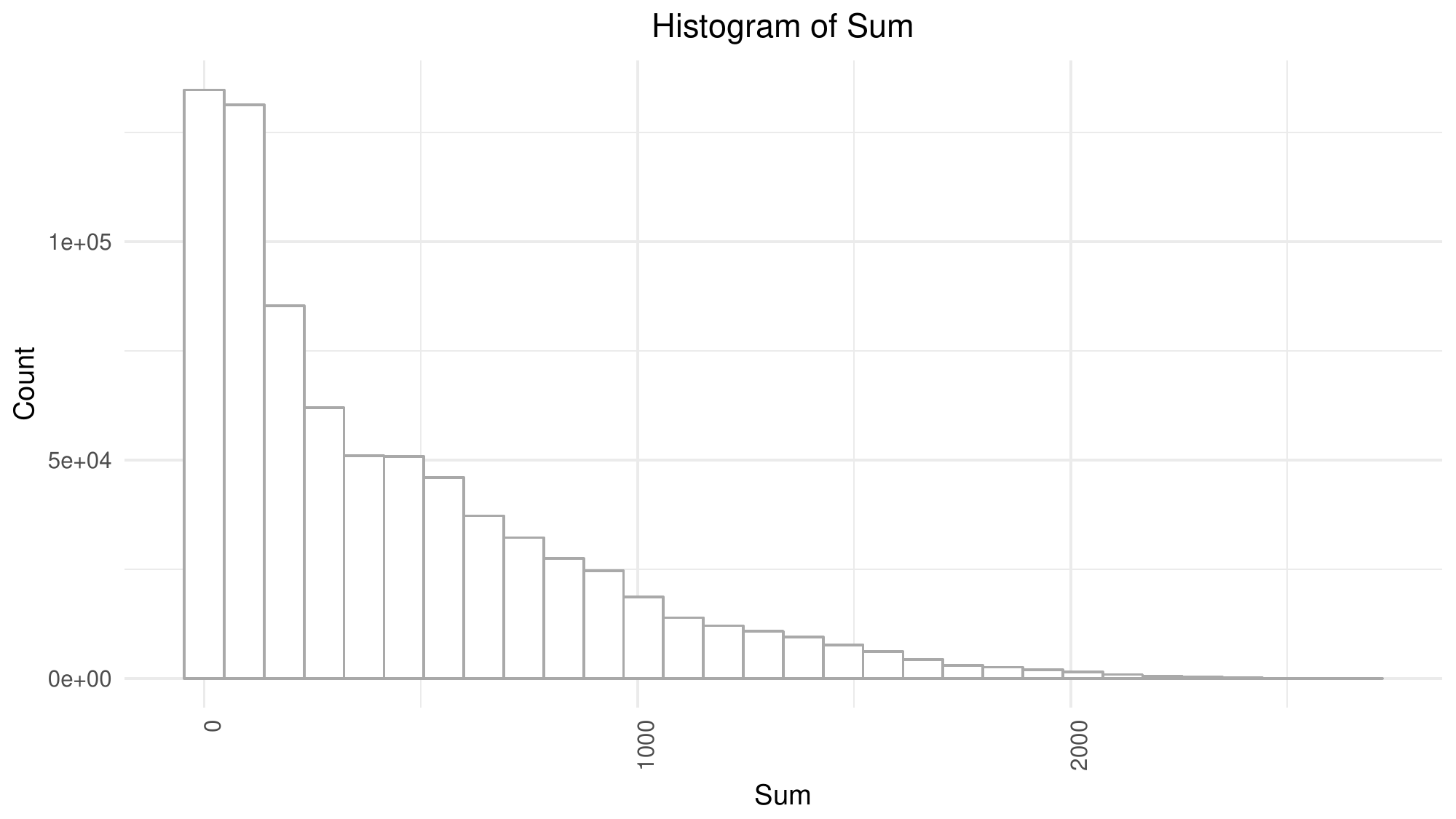}
    \caption{Histogram of the Sum variable, showing a right skewed distribution.}
\end{figure}

The mean per individual site was then calculated. There is a high amount of variation between sites. The mean count ranges from 21.2 for site 87 to 1040 for site 38.
\begin{lstlisting}[basicstyle=\small]
> site_means
 [1]  759.42925  441.98785  448.31918  623.85218  282.13367  764.74474  277.34876  324.37975
 [9]  299.42688  274.73450  415.45149  405.90955  323.46327  594.90033  493.13527  675.71001
[17]  758.92793  621.85340  868.91672  493.19930  323.07675  650.71247  616.00422  593.23522
[25]  483.36278  401.65174  280.81331  227.69805  577.40712  398.69556  416.32734  439.94986
[33]  372.82661  514.34503  431.93221  309.78825  353.07126 1040.16937  396.77779  396.42508
[41]  700.99394  881.75169  583.08680  625.66561  276.10636  664.14990  189.29603  271.97679
[49]   93.87942  314.89299   30.09410  283.12855  857.37885  160.34653  280.71154  276.90954
[57]  871.64668  263.55210  127.70718  460.65924  891.65010  320.14514  328.53539  355.92475
[65]  620.24575  534.23944  963.58047  832.36417  821.75009  170.79412  283.38387  374.61378
[73]  254.41712  583.16942  122.64221  347.20161  174.09591  123.17792  123.64903  249.70587
[81]  137.03924  103.61872  305.51688  118.52460  264.16922  576.14453   21.20215  242.60216
[89]  232.07296  482.56088  172.29681  290.42743  168.53986  399.13894  113.73490  124.29912
\end{lstlisting}

\begin{figure}[H]
\centering
    \includegraphics[width=12cm]{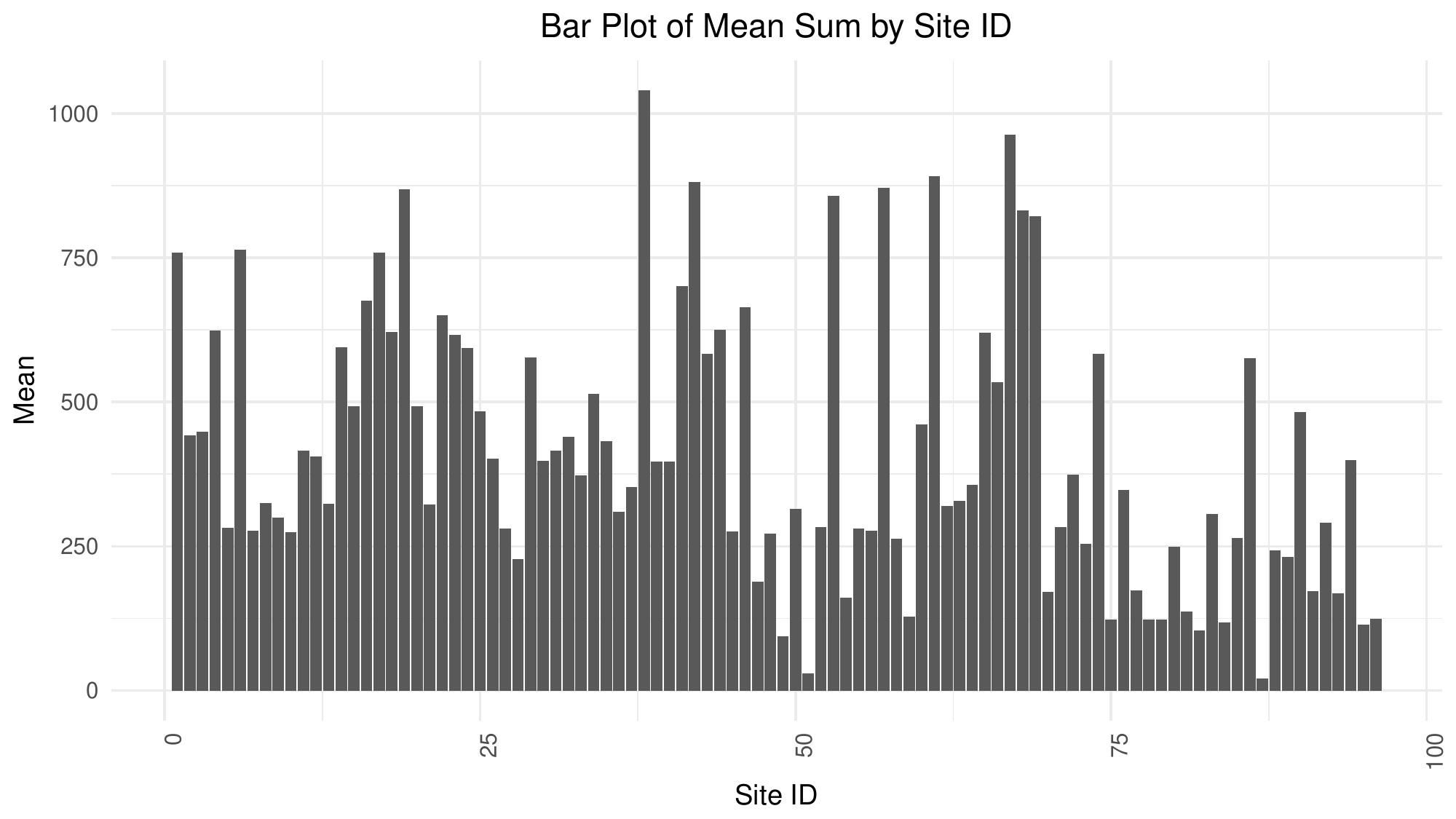}
    \caption{Mean traffic sum per site}
\end{figure}

Plotting the Sum by time shows a slight morning peak followed by a steady rise until the evening peak.

\begin{figure}[H]
\centering
    \includegraphics[width=12cm]{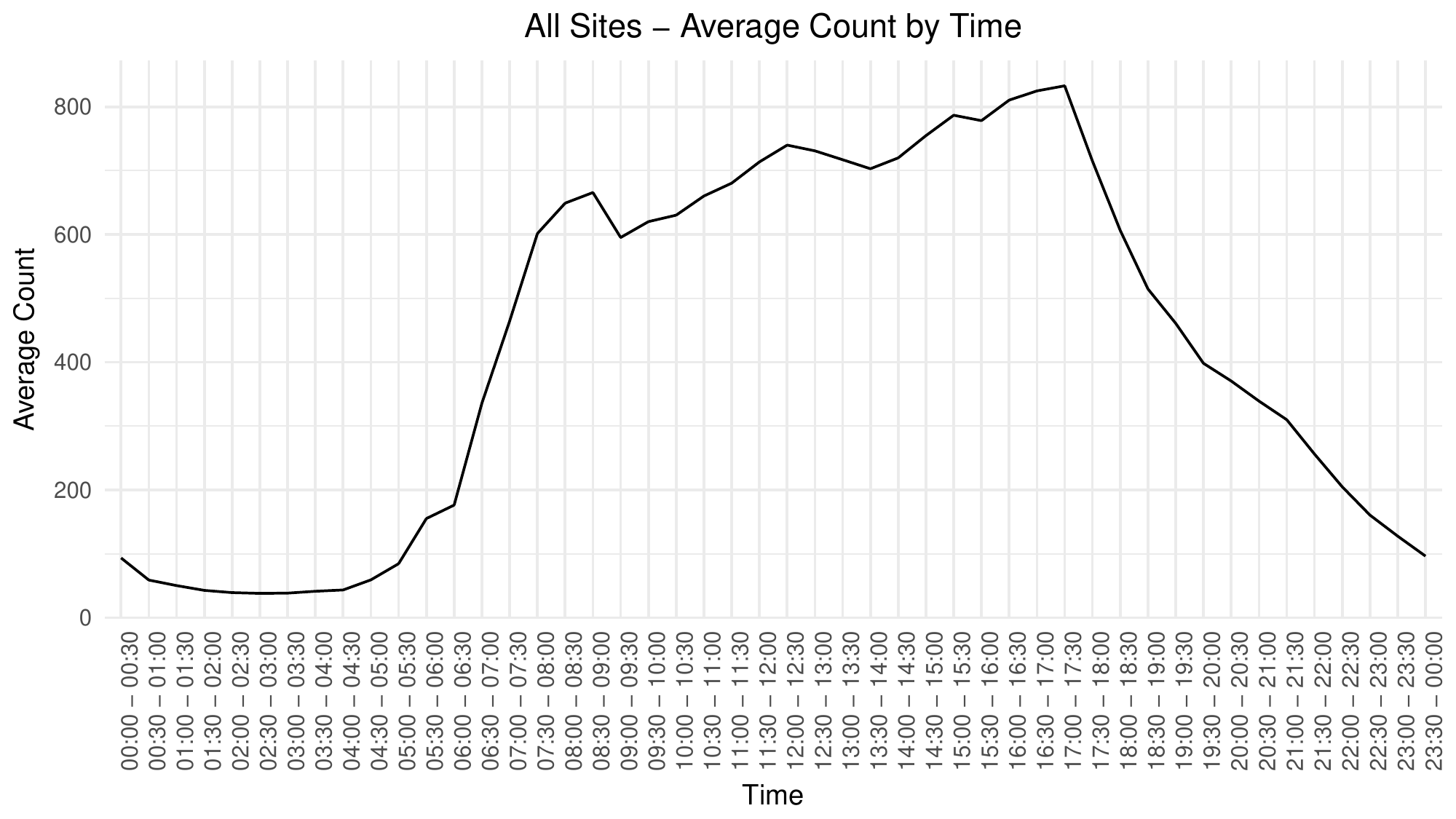}
    \caption{Mean Traffic Count (Sum) across all sites by time.}
\end{figure}

This varies significantly by site, although most sites appear to follow the same general shape over time. The top line is Site 38 - Te Rapa Road/Wairere Drive, which is the only site with a larger lunchtime peak than evening peak. A few sites stand out with an unusual zig-zag shape. These sites record an unusually low count every second half hour. HCC was contacted about this and were unaware of the issue. It is hoped that the INLA prediction will give more realistic predictions for the true traffic count at these sites. It is likely that the prediction error for the sites will be high, so it is useful to note these beforehand.

\begin{figure}[H]
\centering
    \includegraphics[width=14cm]{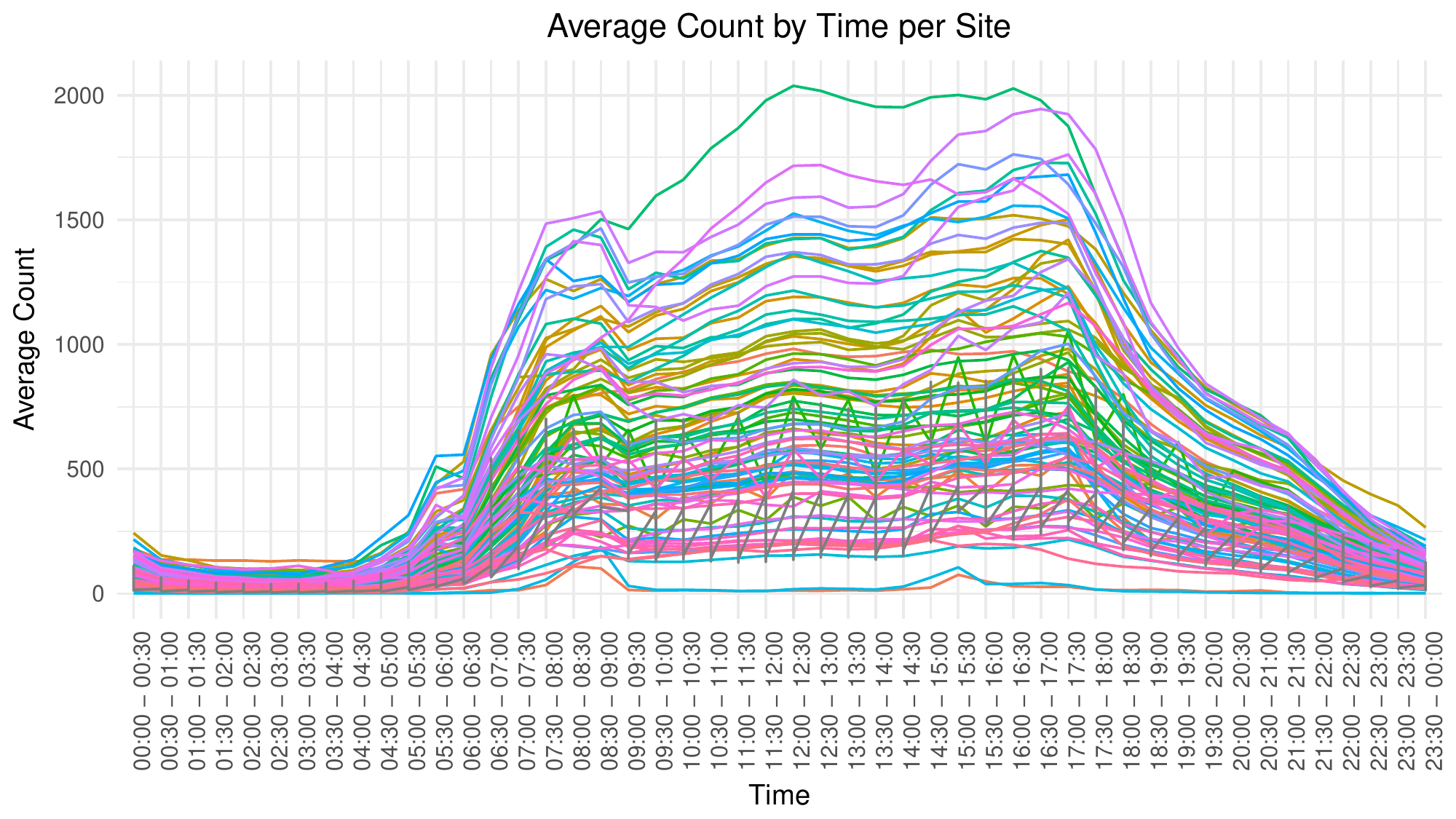}
    \caption{Mean Traffic Count (Sum) across each sites by time.}
\end{figure}

There is a total of 78,241 missing values from 864,658 observations. These values were visualised to understand if certain sites or time periods had an unusual amount of missing values.

\hspace*{-1.5in}
\begin{figure}[H]
    \centerline{\includegraphics[width=20cm]{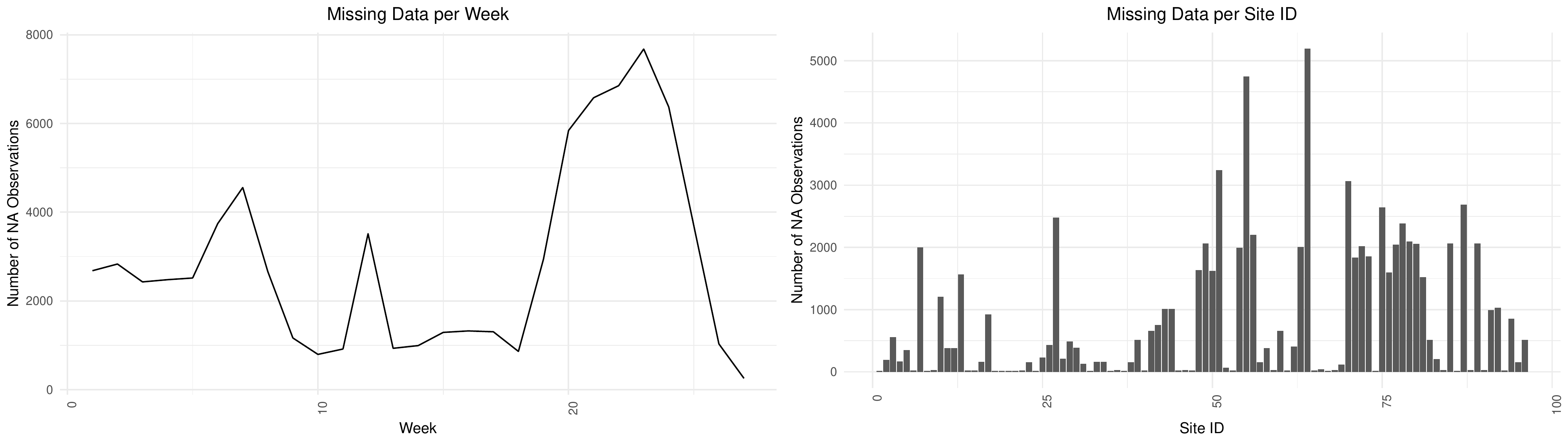}}
    \caption{Left: Missing data per week, from 16 October 2017 to 19 April 2018. Right: Missing data per Site ID.}
\end{figure}

There is a large amount of missing observations from weeks 20 to 24. This period covers both school holidays and school dates. The difference between these periods will be tested, as it is expected that sites near school experience a increase in traffic during school hours. The effect of this will be analysed in the INLA model as a covariate. The chosen time period does not include the Christmas holiday period during which less traffic is expected. The large number of missing data in this time period may have an effect on the predictive accuracy of the model. An additional model was run for the weeks 1-9. There is comparatively a smaller amount of missing data in this period. The difference in Mean Percentage Error was evaluated for days, weeks and sites between the two date ranges. The difference was found to be insignificant. For this reason we chose to subset the data to the date range 15 January 2018 - 13 April 2018. \\

As seen in Figure 5 (right), there are a few sites with an unusually high amount of missing data. When HCC was contacted about this they advised that these sites are minor and as a result their repair is low priority. Often some or all detectors were found to be offline at these sites, although they were still included in the model as there are periods during which they are recording. This data is valuable for the predictions at nearby sites.

\section{Spatio-Temporal Correlation}

The intersections are defined as realisations of the stochastic traffic flow indexed by locations in the space domain $D$. The count $y$ is a random outcome at each location, while the spatial index $s$ varies continuously in the city boundary $D$. The data is represented in space as a collection of observations $y = y(s_{1}),y(s_{n}),...,y(s_{n})$ where $s_{1},...,s_{n}$ indicate the intersection locations.\\

The geographical correlation can be defined with a sparse matrix $n \times n$, where n is the number of observed locations. Each value in the matrix represents the existence of spatial correlation between sites $r$ and $c$, where $r$ represents the location at row index $r$ and $c$ represents the location at column index $c$. If each location only has a few neighbors, in the case of the intersection network, most elements of the matrix will be zero. This greatly reduces the computational complexity of computing the correlation existing in the spatial structure. A sparse matrix was created for the traffic model using GIS. Where two sites were nearby and on the same road, the traffic count at one site is expected to be correlated with the traffic at the second site in the same time interval. The presence of correlation was represented as a '1', with all other cells empty for valid input into the INLA model.\\
\begin{figure}[H]
    \centering
    \includegraphics[width=6cm, height=6cm]{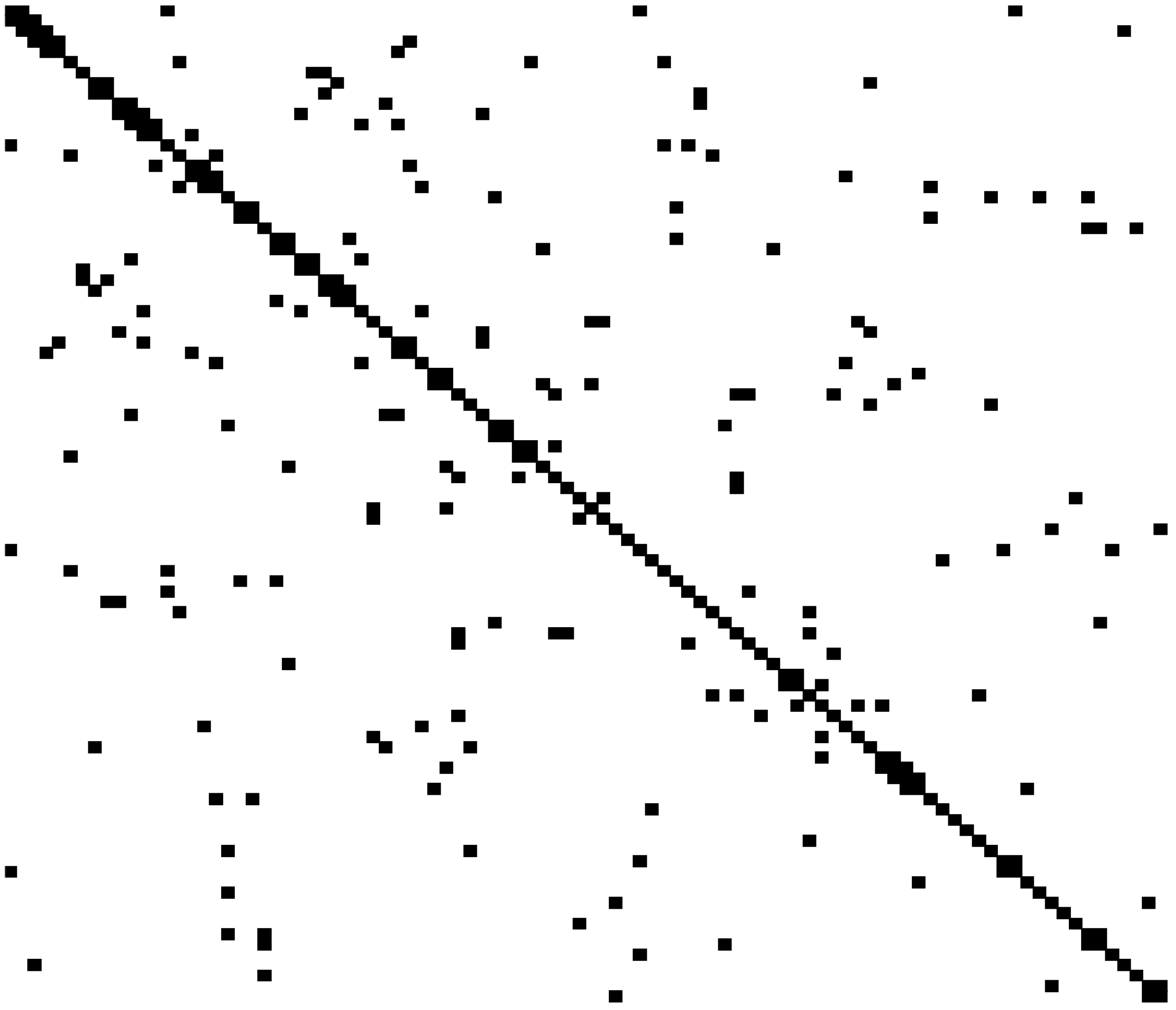}
    \caption{Sparse matrix for the intersections. A black square denotes correlation between the site in row r and column c. Note each index is sequential and not representative of the site number. The sparse matrix was changed for the final two models (13 and 14) which removed three sites.}
\end{figure}

Note: The site numbers were recoded to ID's for input into the INLA model, as INLA expects sequential area ID's. Further reference to sites will used the ID instead of the site number, which may not necessarily be the same. \\

The temporal variation is a Markov process such that $P(X_{n} = j | X_{n-1} = i)$. When using only time as a variable, the traffic count at a particular time value is dependent only on the previous time value. Both the spatial and temporal correlation need to be accounted for in the model. The process indexed by space and time is
\begin{align}
Y(s,t) \equiv {y(s,t),(s,t) \in \box \subset \mathbb{R}^{2} \times \mathbb{R}}
\end{align}
observed at $n$ spatial locations and $T$ time points. For a seasonal model, there is an additional dependence on the same time point in the previous season. A natural definition of a season in the traffic model is a day. For traffic at 8am at a given site is related to the traffic at 7am on the same day, as well as 8am on the previous day. \\

The spatial variable is structured while the time is both structured and unstructured with a random $iid$ effect. The model is defined as
\begin{align}
n_{it} = \beta_{0} + \mu_{i} + \upsilon_{i} + \gamma_{t} + \phi_{t}
\end{align}
where $n_{it}$ is the link function \\
$\beta_{0}$ is the intercept \\ 
$\mu_{i}$ is the structured spatial effect modelled as BYM \\ 
$\upsilon_{i}$ is the unstructured spatial effect modelled as $iid$ \\ 
$\gamma_{t}$ is the structured time effect modelled as seasonal with period p\\ 
and $\phi_{t}$ is the unstructured time effect modelled as $iid$. \\

An interaction effect can be added to the model. This occurs where the change in the response between different spatial areas is dependent on the time period, and vice versa. For example the change in traffic at Site 1 from 5pm-6pm may be different to the same change from 5pm-6pm at Site 2. if no interaction is assumed, this change is assumed to be constant, varying only between time periods and the site. The INLA model allows for multiple interaction types to be specified depending on the structure of the space and time variables. Both space and time are structured in the traffic data. The structure matrix can be written as the Kronecker product of $R_{\delta} = R_{u} \otimes R_{\gamma}$, and has a rank of (T-1)(n-1) for a random walk of order 1, and of (T-2)(n-1) for a random walk of order 2. Although this interaction type is appropriate in the context of the traffic model, we were unable to successfully run the INLA model with this interaction effect. As an alternative we resorted to the Type I interaction, which assumes an interaction between the two unstructured effects time and space. The structure matrix is written as
\begin{align}
R_{\delta} = R_{v} \otimes R_{\phi} = I \otimes I = I
\end{align}
\section{INLA Model}

\subsection{Evaluating Model Performance}

Hamilton City Council have indicated that they are seeking a prediction error less than 10\%. That is for the traffic in one hour at each site, the predicted count is no further than 10\% from the observed count for the majority of total observations. This can be expressed as the average percentage difference between the observed and predicted values. For a given site this is represented as
\begin{align}
MPE_{s,t} = \frac{1}{n}\sum_{i \in s,t}^{}(\frac{abs(y_{i}-\hat{y_{i}})}{y_{i}} \times 100)
\end{align}
Where $MPE_{s,t}$ is the mean percentage deviation of the predicted value from the observed value at site $s$ and time $t$, and $n$ is the total number of observations at site $s$. Equivalently this can be expressed as an MPE for a given time or day across all sites. The overall MPE for all observations is the average of all individual MPE's:
\begin{align}
MPE = \frac{1}{n}\sum_{i=1}^{n}(\frac{abs(y_{i}-\hat{y_{i}})}{y_{i}} \times 100)
\end{align}

The predictive accuracy was evaluated for a full week for each model. Multiple models were tested, beginning with the most basic model using only time, space and no interaction. In addition the performance of each INLA control setting was checked. This was done at the beginning with further models created using the fastest one, provided the accuracy was similar to the rest. Covariates were progressively added, with both interaction and non interaction so that the reduction in effect of the interaction could be evaluated at each step. In addition the effect of missing data was investigated. As seen in Figure 5, the first three months of the data set have a much smaller amount of missing data compared to the last three months. As noted in Section 4.3 the larger amount of missing data in the last three months did not have a noticeable effect on the predictive accuracy of the model.

\subsection{Model 11}

\begin{table}[H]
\centering
\small
\caption{Model 11 Parameters}
\begin{tabular}{|l|l|l|l|l|l|}
\hline
\multicolumn{1}{|l|}{\textbf{Time Aggregation}} & \multicolumn{1}{l|}{\textbf{Interaction}} & \multicolumn{1}{l|}{\textbf{Covariates}} & \multicolumn{1}{l|}{\textbf{Date Range}} & \multicolumn{1}{l|}{\textbf{Prediction Range}} & \multicolumn{1}{l|}{\textbf{Control}} \\ \hline
1 Hour                                & Yes                               & None                            & Weekdays, Jan 15-Apr 13                 & Weekdays, Apr 7 - Apr 13          & Gaussian + EB
\\ \hline             
\end{tabular}
\end{table}

\begin{lstlisting}
model.formula = Y ~ f(ID, model = "bym", graph = H) +
  f(Time, model = "seasonal", season.length = 12) +
  f(Time1, model = "iid") + f(ID.Time, model = "iid")

model = inla(model.formula, family = "poisson", data = data.subset, 
                  control.predictor = list(link = 1, compute = TRUE), 
                  verbose = T,
                  control.inla = list(strategy = "gaussian", int.strategy = "eb"))
\end{lstlisting}

\centering{\Large{\textbf{MPE: 14.25}}}

\begin{table}[H]
\centering
\begin{tabular}{rrrrrr}
  \hline
 & Monday & Tuesday & Wednesday & Thursday & Friday \\ 
  \hline
1 & 16.17 & 14.54 & 14.14 & 13.27 & 13.19 \\ 
   \hline
\end{tabular}
\end{table}

\begin{table}[H]
\centering
\small
\begin{tabular}{rrrrrrrrrrrrr}
  \hline
 & 7-8 & 8-9 & 9-10 & 10-11 & 11-12 & 12-13 & 13-14 & 14-15 & 15-16 & 16-17 & 17-18 & 18-19 \\ 
  \hline
& 21.25 & 16.37 & 9.42 & 12.30 & 14.76 & 14.32 & 13.40 & 9.94 & 12.83 & 10.68 & 14.13 & 21.74 \\  
   \hline
\end{tabular}
\end{table}

\hspace*{-1.5in}
\begin{figure}[H]
    \centerline{\includegraphics[width=20cm]{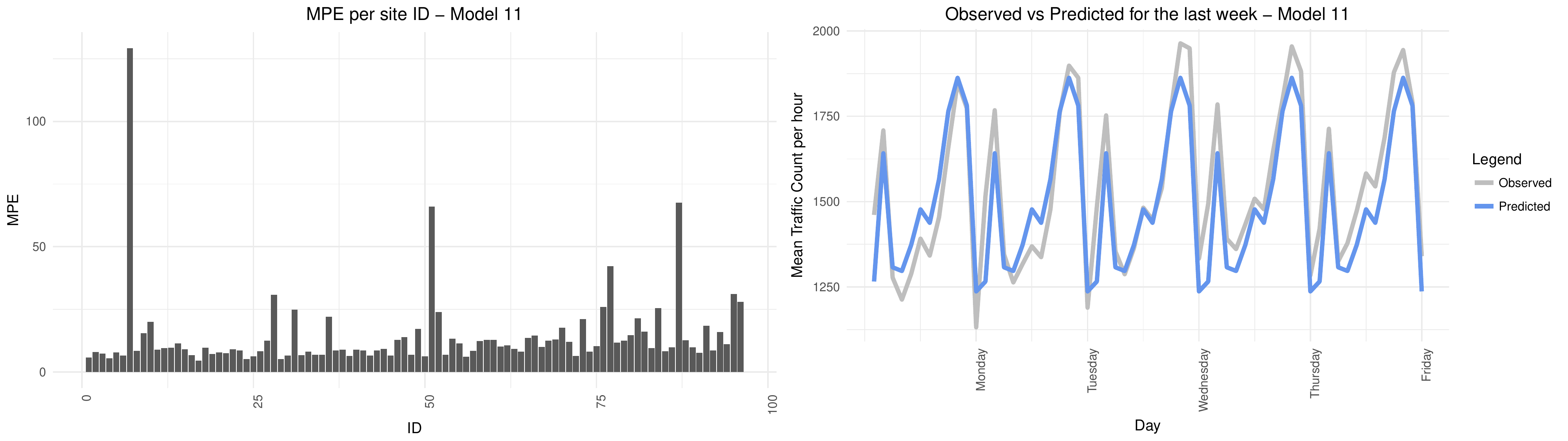}}
    \caption{Left: MPE per site. Right: Observed vs Predicted for last week.}
\end{figure}

\subsection{Model 12}

\begin{table}[H]
\centering
\small
\caption{Model 12 Parameters}
\begin{tabular}{|l|l|l|l|l|l|}
\hline
\multicolumn{1}{|l|}{\textbf{Time Aggregation}} & \multicolumn{1}{l|}{\textbf{Interaction}} & \multicolumn{1}{l|}{\textbf{Covariates}} & \multicolumn{1}{l|}{\textbf{Date Range}} & \multicolumn{1}{l|}{\textbf{Prediction Range}} & \multicolumn{1}{l|}{\textbf{Control}} \\ \hline
1 Hour                                & Yes                               & None                            & Weekends, Jan 15-Apr 13                 & Weekends, Apr 7-Apr 13          & Gaussian + EB
\\ \hline             
\end{tabular}
\end{table}

\begin{lstlisting}
model.formula = Y ~ f(ID, model = "bym", graph = H) +
  f(Time, model = "seasonal", season.length = 12) +
  f(Time1, model = "iid") + f(ID.Time, model = "iid")

model = inla(model.formula, family = "poisson", data = data.subset, 
                  control.predictor = list(link = 1, compute = TRUE), 
                  verbose = T,
                  control.inla = list(strategy = "gaussian", int.strategy = "eb"))
\end{lstlisting}

\centering{\Large{\textbf{MPE: 17.71}}}

\begin{table}[H]
\centering
\begin{tabular}{rrr}
  \hline
 & Saturday & Sunday \\ 
  \hline
1 & 15.28 & 20.18 \\ 
   \hline
\end{tabular}
\end{table}

\begin{table}[H]
\centering
\small
\begin{tabular}{rrrrrrrrrrrrr}
  \hline
 & 7-8 & 8-9 & 9-10 & 10-11 & 11-12 & 12-13 & 13-14 & 14-15 & 15-16 & 16-17 & 17-18 & 18-19 \\ 
  \hline
& 36.67 & 29.59 & 21.04 & 18.17 & 15.58 & 14.06 & 13.36 & 10.76 & 11.51 & 11.50 & 12.79 & 17.77 \\  
   \hline
\end{tabular}
\end{table}

\hspace*{-1.5in}
\begin{figure}[H]
    \centerline{\includegraphics[width=20cm]{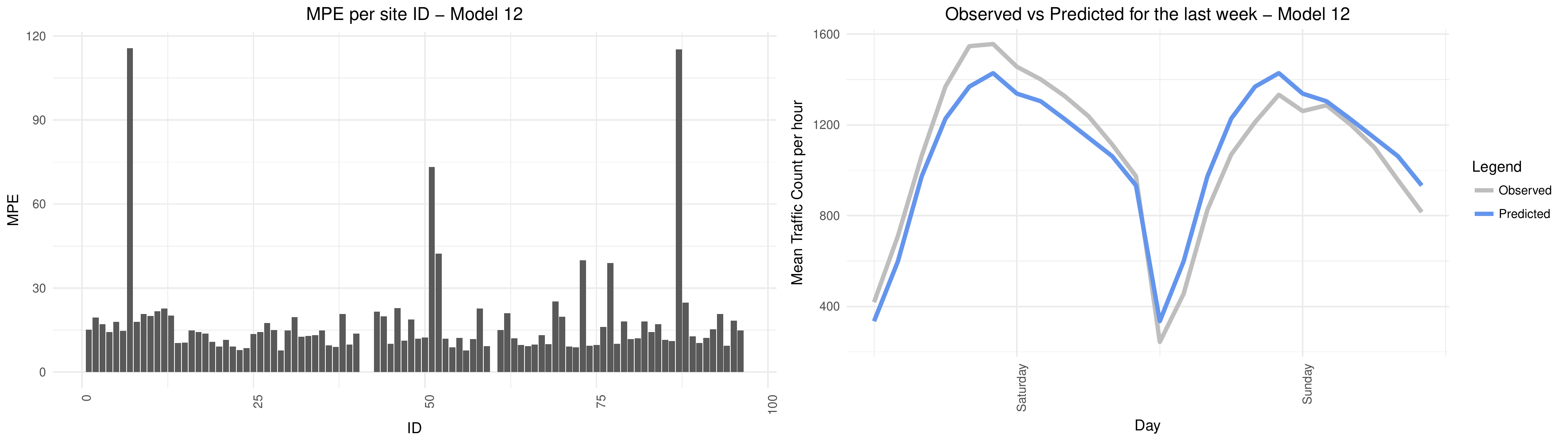}}
    \caption{Left: MPE per site. Right: Observed vs Predicted for last week.}
\end{figure}

\hspace*{-1.5in}
\begin{figure}[H]
    \centerline{\includegraphics[width=20cm]{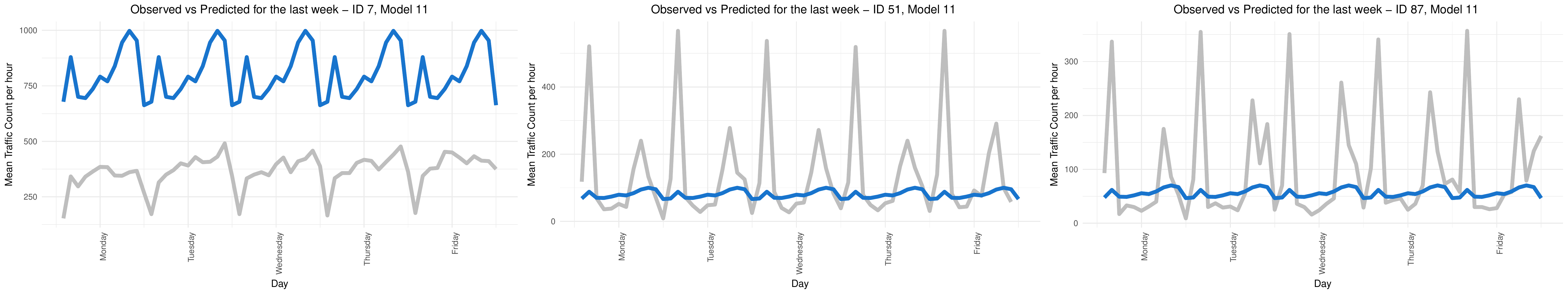}}
    \caption{Left: 3 sites with the highest MPE}
\end{figure}

\begin{flushleft}
There are three obvious outliers in Figure 15 (left). Two of these, IDs 51 and 87, are pedestrian crossings near schools in Rototuna. They experience comparatively very little traffic compared to school drop off and pick up times. The model predicts well for most hours apart from school pick up and drop off times, where it is underpredicting. In addition as seen in the above figure (left), the model is overpredicting for Site 7 by a significant margin. This is assumed to be a fault in the detector recording. The model is likely predicting a more realistic count given the count of nearby sites in each time period. All three sites were deemed by HCC as unimportant for prediction. To investigate the influence of removing these sites on the MPE, two additional models were run for both weekdays and weekends with these three sites removed. The sparse matrix was updated accordingly as described in Section 5.
\end{flushleft}

\subsection{Model 13}

\begin{table}[H]
\centering
\small
\caption{Model 13 Parameters}
\begin{tabular}{|l|l|l|l|l|l|}
\hline
\multicolumn{1}{|l|}{\textbf{Time Aggregation}} & \multicolumn{1}{l|}{\textbf{Interaction}} & \multicolumn{1}{l|}{\textbf{Covariates}} & \multicolumn{1}{l|}{\textbf{Date Range}} & \multicolumn{1}{l|}{\textbf{Prediction Range}} & \multicolumn{1}{l|}{\textbf{Control}} \\ \hline
1 Hour                                & Yes                               & None                            & Weekdays, Jan 15-Apr 13                 & Weekdays, Apr 7-Apr 13          & Gaussian + EB
\\ \hline             
\end{tabular}
\end{table}

\begin{lstlisting}
model.formula = Y ~ f(ID, model = "bym", graph = H) +
  f(Time, model = "seasonal", season.length = 12) +
  f(Time1, model = "iid") + f(ID.Time, model = "iid")

model = inla(model.formula, family = "poisson", data = data.subset, 
                  control.predictor = list(link = 1, compute = TRUE), 
                  verbose = T,
                  control.inla = list(strategy = "gaussian", int.strategy = "eb"))
\end{lstlisting}

\centering{\Large{\textbf{MPE: 11.96}}}

\begin{table}[H]
\centering
\begin{tabular}{rrrrrr}
  \hline
 & Monday & Tuesday & Wednesday & Thursday & Friday \\ 
  \hline
1 & 12.48 & 12.25 & 11.89 & 11.54 & 11.64 \\ 
   \hline
\end{tabular}
\end{table}

\begin{table}[H]
\centering
\small
\begin{tabular}{rrrrrrrrrrrrr}
  \hline
 & 7-8 & 8-9 & 9-10 & 10-11 & 11-12 & 12-13 & 13-14 & 14-15 & 15-16 & 16-17 & 17-18 & 18-19 \\ 
  \hline
1 & 18.05 & 14.64 & 7.69 & 10.43 & 11.84 & 11.91 & 11.71 & 8.02 & 10.05 & 8.92 & 12.58 & 17.66 \\ 
   \hline
\end{tabular}
\end{table}

\hspace*{-1.5in}
\begin{figure}[H]
    \centerline{\includegraphics[width=20cm]{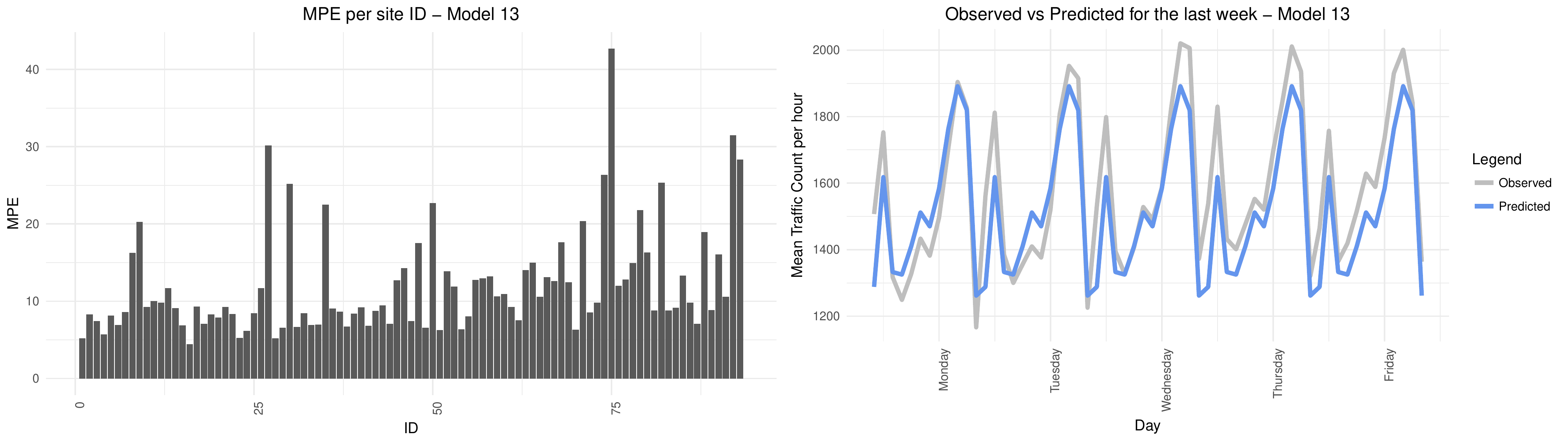}}
    \caption{Left: MPE per site. Right: Observed vs Predicted for last week.}
\end{figure}

\begin{table}[H]
\centering
\begin{tabular}{rrrrrrrrrrrrrrrrrrrrrrrrrrrrrrrrrrrrrrrrrrrrrrrrrrrrrrrrrrrrrrrrrrrrrrrrrrrrrrrrrrrrrrrrrrrrr}
  \hline
 1 & 2 & 3 & 4 & 5 & 6 & 7 & 8 & 9 & 10 & 11 & 12 & 13 & 14 & 15 \\
  \hline
5.22 & 8.30 & 7.42 & 5.73 & 8.16 & 6.94 & 8.62 & 16.28 & 20.27 & 9.29 & 10.01 & 9.85 & 11.72 & 9.13 & 6.88 \\\\
   \hline
    16 & 17 & 18 & 19 & 20 & 21 & 22 & 23 & 24 & 25 & 26 & 27 & 28 & 29 & 30 \\ 
    \hline
     4.46 & 9.34 & 7.07 & 8.29 & 7.89 & 9.24 & 8.34 & 5.25 & 6.18 & 8.46 & 11.71 & 30.16 & 5.20 & 6.56 & 25.18 \\\\ 
     \hline
     31 & 32 & 33 & 34 & 35 & 36 & 37 & 38 & 39 & 40 & 41 & 42 & 43 & 44 & 45 \\
     \hline
      6.67 & 8.46 & 6.93 & 7.00 & 22.51 & 9.06 & 8.66 & 6.74 & 8.41 & 9.21 & 6.83 & 8.77 & 9.45 & 7.10 & 12.69 \\\\
      \hline
      46 & 47 & 48 & 49 & 50 & 51 & 52 & 53 & 54 & 55 & 56 & 57 & 58 & 59 & 60 \\
      \hline
       14.27 & 7.46 & 17.51 & 6.57 & 22.73 & 6.27 & 13.86 & 11.91 & 6.35 & 8.05 & 12.74 & 12.98 & 13.22 & 10.64 & 10.96 \\\\
       \hline 
       61 & 62 & 63 & 64 & 65 & 66 & 67 & 68 & 69 & 70 & 71 & 72 & 73 & 74 & 75 \\
       \hline
        9.24 & 7.52 & 14.01 & 15.00 & 10.58 & 13.10 & 12.62 & 17.63 & 12.47 & 6.30 & 20.37 & 8.56 & 9.83 & 26.35 & 42.69 \\\\
        \hline
         76 & 77 & 78 & 79 & 80 & 81 & 82 & 83 & 84 & 85 & 86 & 87 & 88 & 89 & 90 \\
         \hline 
         12.01 & 12.84 & 14.96 & 21.80 & 16.29 & 8.82 & 25.34 & 8.80 & 9.15 & 13.31 & 9.83 & 7.11 & 18.96 & 8.86 & 16.08 \\\\
         \hline
          91 & 92 & 93 \\ 
          \hline
           10.60 & 31.50 & 28.35
\end{tabular}
\caption{MPE per site}
\end{table}

\begin{table}[H]
\centering
\begin{tabular}{rrrrrr}
  \hline
 & Monday & Tuesday & Wednesday & Thursday & Friday \\ 
  \hline
7-8 & 17.34 & 18.63 & 19.26 & 18.47 & 16.54 \\ 
  8-9 & 14.25 & 14.59 & 14.77 & 15.75 & 13.85 \\ 
  9-10 & 6.19 & 7.57 & 7.69 & 9.47 & 7.55 \\ 
  10-11 & 12.43 & 11.29 & 9.06 & 9.63 & 9.74 \\ 
  11-12 & 13.95 & 11.55 & 12.10 & 10.50 & 11.11 \\ 
  12-13 & 13.36 & 15.31 & 12.30 & 9.13 & 9.45 \\ 
  13-14 & 13.59 & 14.44 & 10.43 & 9.95 & 10.15 \\ 
  14-15 & 8.85 & 8.40 & 5.75 & 8.33 & 8.75 \\ 
  15-16 & 10.19 & 7.96 & 9.37 & 9.50 & 13.21 \\ 
  16-17 & 8.89 & 7.80 & 8.72 & 9.36 & 9.84 \\ 
  17-18 & 12.14 & 12.74 & 13.55 & 12.04 & 12.42 \\ 
  18-19 & 18.58 & 16.67 & 19.61 & 16.40 & 17.06 \\ 

   \hline
\end{tabular}
\caption{MPE per day and time. Note the time is in 24 hour format.}
\end{table}

\subsection{Model 14}

\begin{table}[H]
\centering
\small
\caption{Model 14 Parameters}
\begin{tabular}{|l|l|l|l|l|l|}
\hline
\multicolumn{1}{|l|}{\textbf{Time Aggregation}} & \multicolumn{1}{l|}{\textbf{Interaction}} & \multicolumn{1}{l|}{\textbf{Covariates}} & \multicolumn{1}{l|}{\textbf{Date Range}} & \multicolumn{1}{l|}{\textbf{Prediction Range}} & \multicolumn{1}{l|}{\textbf{Control}} \\ \hline
1 Hour                                & Yes                               & None                            & Weekends, Jan 15-Apr 13                 & Weekends, Apr 7-Apr 13          & Gaussian + EB
\\ \hline             
\end{tabular}
\end{table}

\begin{lstlisting}
model.formula = Y ~ f(ID, model = "bym", graph = H) +
  f(Time, model = "seasonal", season.length = 12) +
  f(Time1, model = "iid") + f(ID.Time, model = "iid")

model = inla(model.formula, family = "poisson", data = data.subset, 
                  control.predictor = list(link = 1, compute = TRUE), 
                  verbose = T,
                  control.inla = list(strategy = "gaussian", int.strategy = "eb"))
\end{lstlisting}

\centering{\Large{\textbf{MPE: 15.06}}}

\begin{table}[H]
\centering
\begin{tabular}{rrr}
  \hline
 & Saturday & Sunday \\ 
  \hline
1 & 12.67 & 17.45 \\ 
   \hline
\end{tabular}
\end{table}

\begin{table}[H]
\centering
\small
\begin{tabular}{rrrrrrrrrrrrr}
  \hline
 & 7-8 & 8-9 & 9-10 & 10-11 & 11-12 & 12-13 & 13-14 & 14-15 & 15-16 & 16-17 & 17-18 & 18-19 \\ 
  \hline
& 34.45 & 25.83 & 15.24 & 14.38 & 13.79 & 10.86 & 10.62 & 9.21 & 8.99 & 9.54 & 10.77 & 17.05 \\
   \hline
\end{tabular}
\end{table}

\hspace*{-1.5in}
\begin{figure}[H]
    \centerline{\includegraphics[width=20cm]{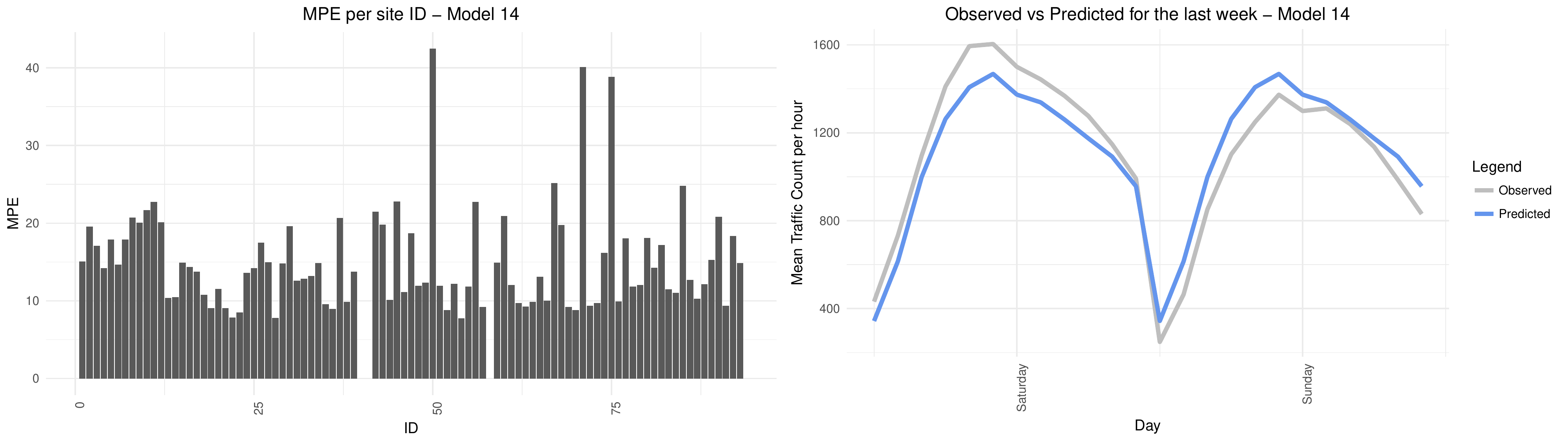}}
    \caption{Left: MPE per site. Right: Observed vs Predicted for last week.}
\end{figure}

\begin{table}[H]
\centering
\begin{tabular}{rrrrrrrrrrrrrrrrrrrrrrrrrrrrrrrrrrrrrrrrrrrrrrrrrrrrrrrrrrrrrrrrrrrrrrrrrrrrrrrrrrrrrrrrrrrrrr}
  \hline
1 & 2 & 3 & 4 & 5 & 6 & 7 & 8 & 9 & 10 & 11 & 12 & 13 & 14 & 15 \\
  \hline
1 15.10 & 19.55 & 17.11 & 14.23 & 17.89 & 14.68 & 17.89 & 20.72 & 20.06 & 21.70 & 22.75 & 20.13 & 10.39 & 10.46 & 14.91 \\\\
   \hline
   16 & 17 & 18 & 19 & 20 & 21 & 22 & 23 & 24 & 25 & 26 & 27 & 28 & 29 & 30 \\
   \hline
   14.35 & 13.79 & 10.77 & 9.09 & 11.52 & 9.06 & 7.87 & 8.49 & 13.62 & 14.24 & 17.52 & 14.99 & 7.80 & 14.81 & 19.61 \\\\ 
   \hline
31 & 32 & 33 & 34 & 35 & 36 & 37 & 38 & 39 & 40 & 41 & 42 & 43 & 44 & 45 \\  
\hline
12.62 & 12.84 & 13.23 & 14.87 & 9.58 & 8.97 & 20.69 & 9.86 & 13.77 &  &  & 21.51 & 19.83 & 10.10 & 22.81 \\\\
 \hline
46 & 47 & 48 & 49 & 50 & 51 & 52 & 53 & 54 & 55 & 56 & 57 & 58 & 59 & 60 \\
\hline
 11.14 & 18.70 & 11.94 & 12.34 & 42.48 & 11.97 & 8.81 & 12.19 & 7.73 & 11.82 & 22.73 & 9.23 &  & 14.94 & 20.91 \\\\
 \hline
 61 & 62 & 63 & 64 & 65 & 66 & 67 & 68 & 69 & 70 & 71 & 72 & 73 & 74 & 75 \\
 \hline
 12.06 & 9.70 & 9.27 & 9.86 & 13.08 & 10.01 & 25.15 & 19.77 & 9.20 & 8.80 & 40.09 & 9.39 & 9.73 & 16.16 & 38.85 \\\\
 \hline
 76 & 77 & 78 & 79 & 80 & 81 & 82 & 83 & 84 & 85 & 86 & 87 & 88 & 89 & 90 \\
 \hline
 9.94 & 18.03 & 11.84 & 12.02 & 18.12 & 14.28 & 17.18 & 11.50 & 11.03 & 24.82 & 12.71 & 10.30 & 12.15 & 15.26 & 20.84 \\\\
 \hline
 91 & 92 & 93  \\
 \hline
9.37 & 18.36 & 14.85 \\
 \hline
\end{tabular}
\caption{MPE per site}
\end{table}

\begin{table}[H]
\centering
\begin{tabular}{rrr}
  \hline
 & Saturday & Sunday \\ 
  \hline
7-8 & 19.93 & 48.98 \\ 
  8-9 & 16.05 & 35.62 \\ 
  9-10 & 11.19 & 19.29 \\ 
  10-11 & 13.93 & 14.83 \\ 
  11-12 & 13.52 & 14.06 \\ 
  12-13 & 11.61 & 10.11 \\ 
  13-14 & 11.29 & 9.96 \\ 
  14-15 & 11.18 & 7.24 \\ 
  15-16 & 10.30 & 7.68 \\ 
  16-17 & 10.43 & 8.65 \\ 
  17-18 & 9.02 & 12.53 \\ 
  18-19 & 13.63 & 20.47 \\ 
   \hline
\end{tabular}
\caption{MPE per day and time. Note the time is in 24 hour format.}
\end{table}

\begin{flushleft}
Removing site IDs 7, 51 and 87 resulted in a significant reduction across the MPE for each day and time. As seen in the MPE per site tables (Table 17 and 20), the MPE for most sites is close to or below the desired 10\% error. For the weekdays model, a few sites remain with 10-30\% error. IDs 92 and 93 are relatively high, and ID 75 has the highest MPE at 42.69\%. In the weekends model, IDs 92 and 93 are performing better while ID 75 is still high. The two models lead to different MPE's for each ID. In some cases ID's with a low MPE in the weekday model have a high MPE in the weekend model, and vice versa. \\

In the weekday model, the MPE 9am-10am across all days is low. This is also true for 2pm-6pm. There is not a large difference between days. In the weekends model, the MPE during Sunday from 7am-9am and 6pm-7pm is high, while 12pm-5pm is low. The MPE for Saturday is relatively constant, but 7am-9am is high, consistent with the rest of the week. Overall the model is underpredicting for most times in the weekdays and Saturday, but overpredicting for Sunday. \newline

The final model is defined as
\begin{align}
n_{it} = \beta_{0} + \mu_{i} + \upsilon_{i} + \gamma_{t} + \phi_{t}
\end{align}
where $n_{it}$ is the link function \\
$\beta_{0}$ is the intercept \\ 
$\mu_{i}$ is the structured spatial effect modelled as BYM \\ 
$\upsilon_{i}$ is the unstructured spatial effect modelled as $iid$ \\ 
$\gamma_{t}$ is the structured time effect modelled as seasonal with period 12\\ 
and $\phi_{t}$ is the unstructured time effect modelled as $iid$. \\
\end{flushleft}

\raggedright
\section{Comparison to prediction using the prior mean}

In the Bayesian context, a prior can be defined for the prediction of traffic count at a given site, time and day. For equivalent comparison to the prediction given by the INLA model, this was taken as the mean of the previous seven weeks. Given that the predictors follow a fixed set of values, it is expected that using solely the prior mean will give an accurate prediction. \newline

For all observations at 7am on Monday, the mean was taken for all 7am Monday observations for the first 8 weeks of our date range and compared to the observed value at 7am Monday for the final week. This was compared to the prediction from Model 11. The MPE for the INLA model was only slightly smaller than the prediction by taking the mean of the previous observations for the same time and day.

\hspace*{-1.5in}
\begin{figure}[H]
    \centerline{\includegraphics[width=14cm]{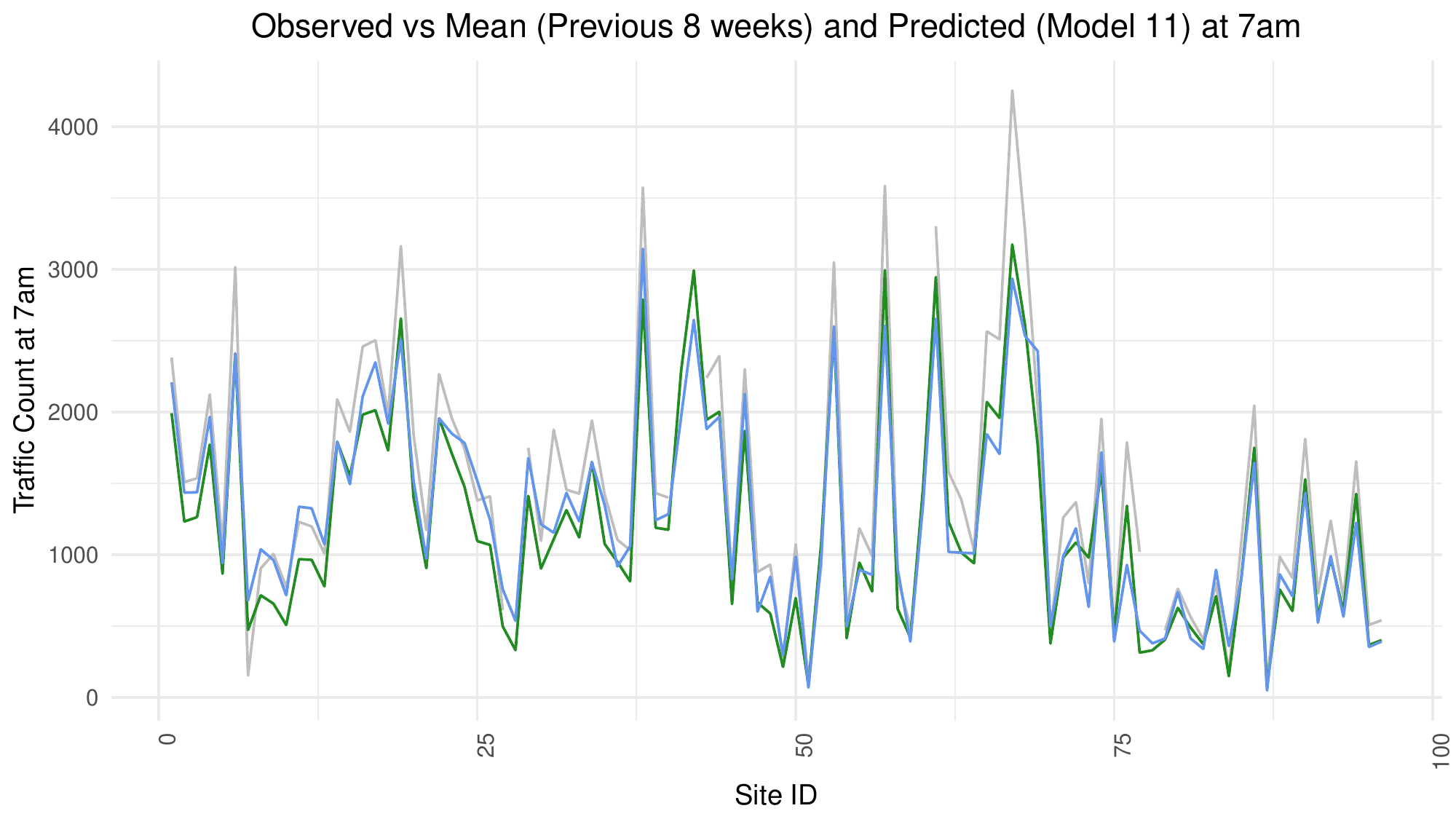}}
    \caption{Observed vs Mean and Predicted for 7am (all sites)}
\end{figure}

\begin{lstlisting}
mean(meanPE, na.rm = T) = 22.30767
mean(predPE, na.rm = T) = 21.18661
\end{lstlisting}

The above comparison was extended to all times on Monday to see if this remains constant for other times from 7am-7pm. This is shown below for a sample of sites.

\hspace*{-1.5in}
\begin{figure}[H]
    \centerline{\includegraphics[width=16cm]{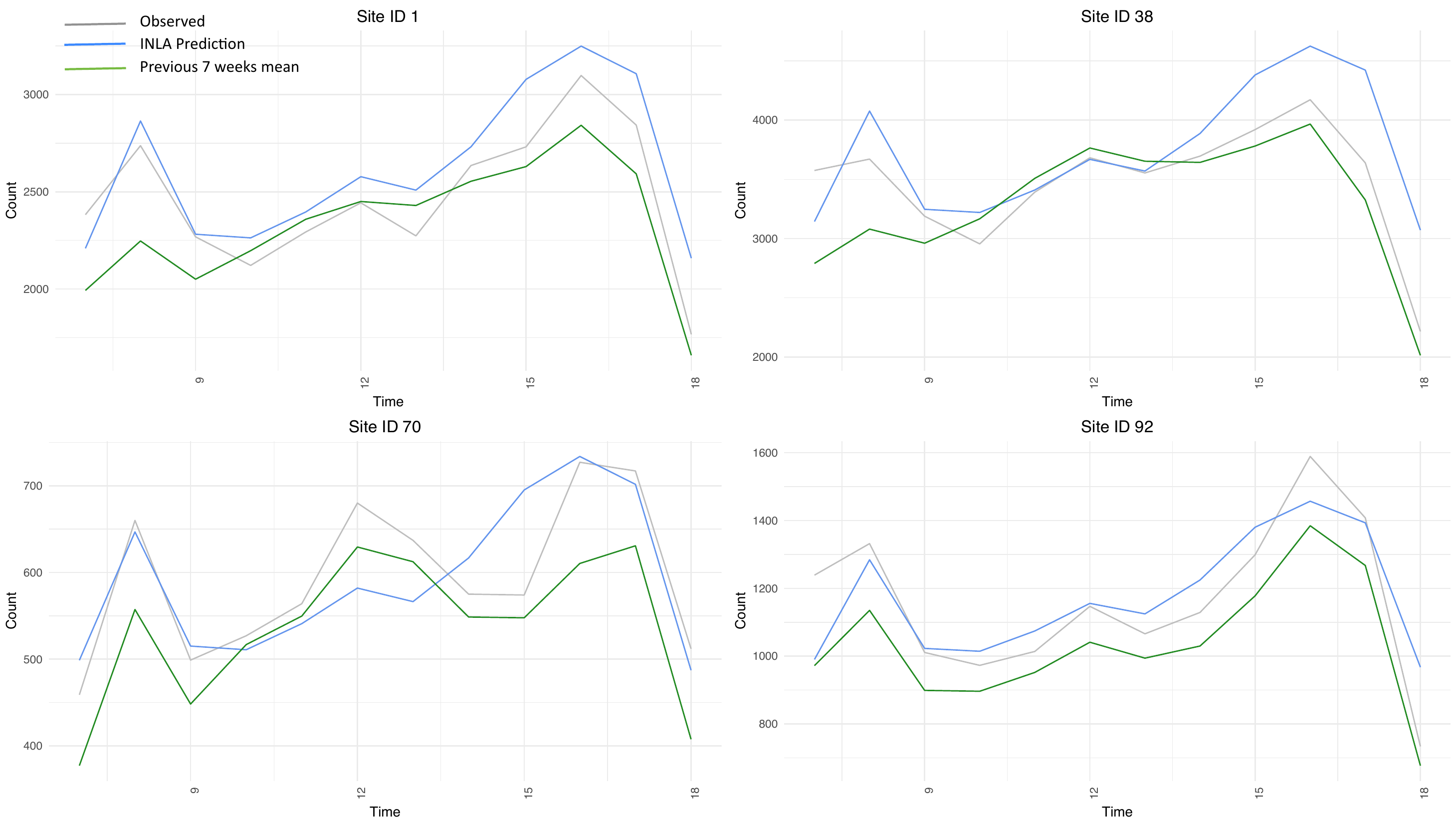}}
    \caption{Observed vs Mean of prior 7 weeks vs Predicted for Monday, Site IDs 1, 38, 70 and 92}
\end{figure}

\begin{lstlisting}
mean(meanPE, na.rm = T) = 11.6286
mean(predPE, na.rm = T) = 16.17448
\end{lstlisting}

In the above, meanPE is the MPE between the prior 7 week mean and the observed count for the same site, time and day. predPE is the MPE between the INLA prediction using model 11 and the observed count for the same site, time and day. The MPE using the prior mean was 11.63\%, compared to 16.17\% for the INLA prediction. \newline

The INLA prediction seems to better predict the traffic count at peak times, although the mean prediction more closely follows the pattern from 7am-7pm. This suggests that the pattern of the traffic is relatively similar across the same site, hour and day. Although this may be affected by covariates such as schools, weather, roadworks and major events, in most cases the mean prediction is sufficient. When the nature of the data is not as consistent or predictable, INLA will likely be more accurate. The mean prediction does not take into account covariates, and therefore when these are highly influential the INLA prediction is likely to outperform the mean prediction. \newline

Perhaps using this process as a prior for the INLA model will result in a more accurate model overall. The prior in this model will act as a baseline, with the further complexity of the spatio-temporal structure and the INLA algorithm providing additional accuracy for the predicted time period. INLA provides a robust structure for defining a prior, although it was not used in the final model.

\begin{table}[H]
\centering
\begin{tabular}{rrrrrr}
  \hline
 & ActualY & pred & mean & meanPE & predPE \\ 
  \hline
1 & 2382 & 2208.88 & 1992.42 & 16.36 & 7.27 \\ 
  2 & 1509 & 1435.24 & 1232.83 & 18.30 & 4.89 \\ 
  3 & 1536 & 1437.64 & 1263.27 & 17.76 & 6.40 \\ 
  4 & 2124 & 1966.71 & 1772.83 & 16.53 & 7.41 \\ 
  5 & 1062 & 941.12 & 866.36 & 18.42 & 11.38 \\ 
  6 & 3017 & 2411.92 & 2407.83 & 20.19 & 20.06 \\ 
  7 & 153 & 678.36 & 472.29 & 208.68 & 343.37 \\ 
  8 & 904 & 1038.13 & 715.25 & 20.88 & 14.84 \\ 
  9 & 1005 & 962.24 & 656.67 & 34.66 & 4.25 \\ 
  10 & 772 & 714.95 & 506.50 & 34.39 & 7.39 \\ 
  11 & 1230 & 1336.73 & 968.67 & 21.25 & 8.68 \\ 
  12 & 1198 & 1325.13 & 964.25 & 19.51 & 10.61 \\ 
  13 & 1007 & 1072.79 & 777.75 & 22.77 & 6.53 \\ 
  14 & 2090 & 1793.84 & 1792.08 & 14.25 & 14.17 \\ 
  15 & 1861 & 1494.27 & 1552.08 & 16.60 & 19.71 \\ 
  16 & 2459 & 2109.15 & 1981.83 & 19.40 & 14.23 \\ 
  17 & 2504 & 2349.14 & 2013.18 & 19.60 & 6.18 \\ 
  18 & 1984 & 1918.44 & 1730.58 & 12.77 & 3.30 \\ 
  19 & 3164 & 2508.33 & 2656.75 & 16.03 & 20.72 \\ 
  20 & 1839 & 1514.82 & 1397.50 & 24.01 & 17.63 \\ 
   \hline
\end{tabular}
   \caption{Comparison between the observed count, prior mean, and INLA prediction for the first twenty site ID's at 7am Monday April 9}
\end{table}

As seen in Figure 27 below, INLA appears to provide more accurate imputations where a response was provided for time periods around the imputed point. Most detectors were down at Site 17 during a certain hourly time period, and therefore the count was recorded as significantly lower than the previous and following hours. The yellow line shows the INLA prediction. The prediction for the faulty recording period appears to be much a realistic estimate of the expected count. As the INLA imputation takes into account the spatio-temporal structure in addition to the historical and following data, it is likely a more precise estimate than the prior mean method. \newline

\hspace*{-1.5in}
\begin{figure}[H]
    \centerline{\includegraphics[width=16cm]{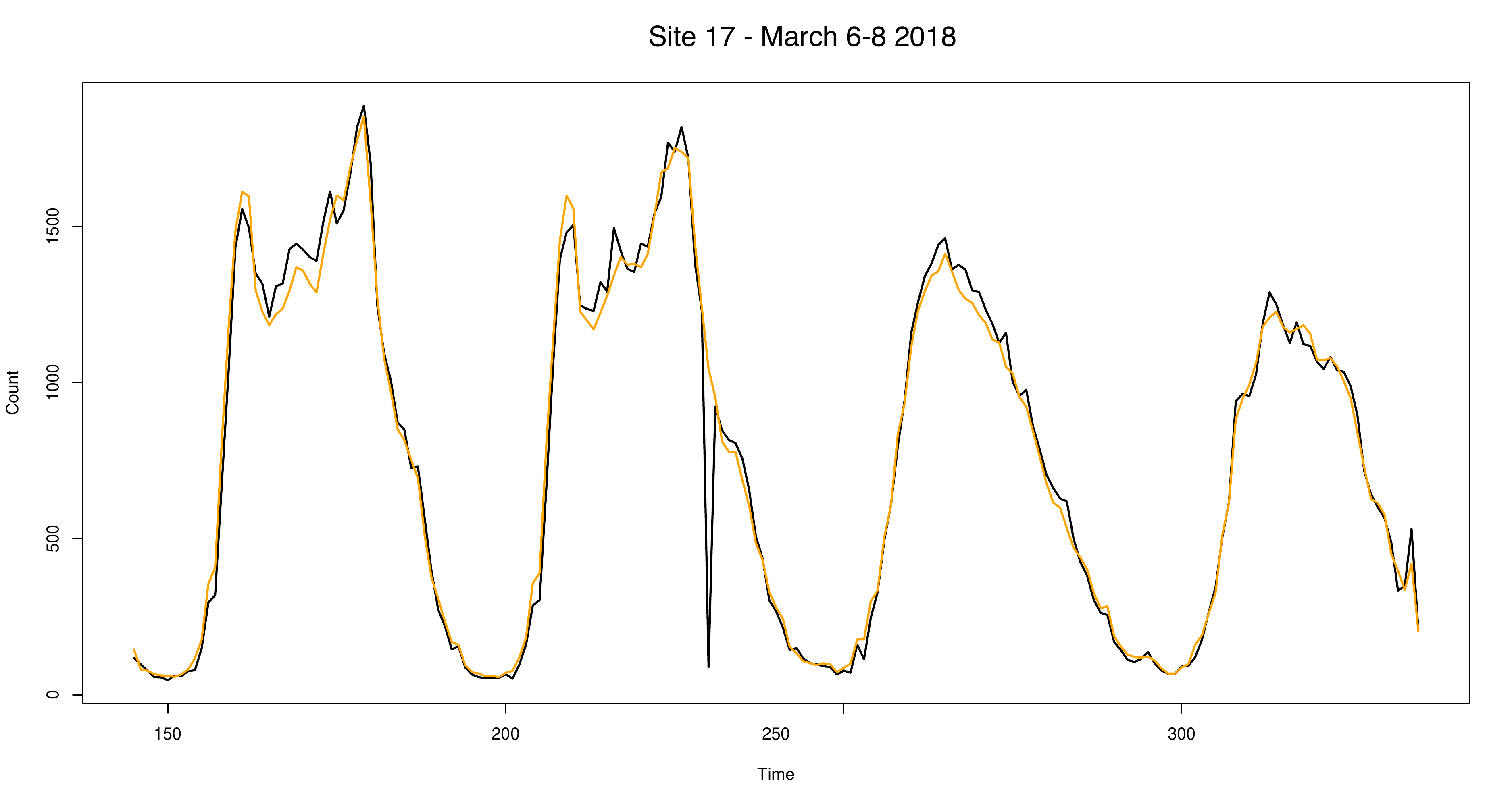}}
    \caption{A detector error imputed by the INLA model. The yellow line is the predicted count by the INLA model.}
\end{figure}

\raggedright
\section{Future Considerations}

Further improvements to the model can be made by investigating the underperforming sites, times and days. As seen in the final model for both weekdays and weekends, there are particular sites with a high MPE. These should be investigated further for the reason. The morning and evening periods have the highest MPE. There could be various reasons for this, such as certain sites experiencing a disproportionate amount of traffic in the morning and evening peak compared to hours outside of peak times. This was seen in the two sites at Rototuna. Removing these sites in the final two models resulted in a significant gain in prediction accuracy, although additions to the model formula such as a structured Type IV interaction could compensate for this disparity in traffic at peak times compared to other sites. \newline

The Type IV interaction should result in a gain in prediction accuracy, given that both time and space are structured in the model. In addition to this, an interaction effect could be explored between day and time or weekend/weekday and time to better fit the difference in shape between days. \newline

Using a fixed effect for schools did not result in a significant gain in prediction accuracy, although an interaction effect could lead to a different result. Different methods of inputting the school covariate into the INLA model should be investigated. Additional covariates such as roadworks, weather and major events could be explored. A proportion of the prediction error in our final model could be explained by these covariates. For example, the traffic during the Sevens rugby tournament would be expected to be higher in areas near FMG Stadium. We are likely seeing underprediction for sites in this area during this time, as our model included the dates of the tournament. \newline

A significant advantage of INLA is the ability to use SPDE models. This extends the prediction range to a continuous domain, which would be particularly useful in the context of traffic prediction. With the fast growing nature of Hamilton, new roads are regularly constructed and existing intersections are modified. The ability to attain an accurate prediction of the traffic count at an intersection not currently controlled by SCATS is extremely valuable for planning purposes. The base form of the INLA SPDE model extends predictions over a continuous domain with no defined boundaries. [2]. This is useful for applications such as prediction over a plot of land or rainfall across a city. In these examples there are no constraints on where a prediction can be made in the spatial area. However in the context of traffic prediction, boundaries need to be defined as roads in the domain of the HCC road network. The paper 'Non-stationery Gaussian models with physical barriers (Bakka, Vanhatalo, Illian, Simpson \& Rue, 2016) is an ongoing exploration of the use of these models in INLA. The addition of this feature to the current model has the potential to serve as an important tool for future planning in Hamilton.

\section{Conclusion}

There is a great benefit in having a model that can accurately predict the traffic counts for certain sites, at a given time and/or day. The information about the traffic dispersion and traffic counts, is essential for future planning regarding the infrastructure of the city. In this report, the two final INLA models separating weekdays and weekends performed well for prediction. We have found that for this data using the mean for the same site, hour and day from several weeks prior was almost equally accurate. Although we could not identify any covariates that are significant in our model, this avenue can be explored more in an attempt to get the overall MPE of  models to less than 10\%. The true utility of INLA will be in its ability to predict in yet unobserved locations. By using a more complex barrier model with additional interactions and covariates, predictions could be made at locations without SCATS control. For example, there are certain intersections in Hamilton changing from roundabouts or crossroads to signalised intersections. As traffic data is not currently collected at these locations, INLA could provide a cost effective method for understanding the traffic flow. \newline

As with all regression models, a key advantage to using an historical mean is the ability to account for unseen or extreme predictor values. If a historical mean is used, a prediction can only be obtained using a specific set of predictor values which have been previously observed. In the case of the traffic models, the key predictors are day, time and location. The values of each of these fall within an expected range with no chance of outliers. If the predictors do not meet this criteria, using a model such as INLA will be a much more effective method than using an historical mean. \newline

While the prior mean and macro method were simpler and provided a sufficiently accurate estimate, INLA was effective in modelling the complex spatio-temporal structure. It provided sufficiently precise predictions, and where the existing data was missing or unusual the model was able to use the data from neighboring locations and times to impute the count. The algorithm's flexibility will likely result in opportunities to use it at HCC where current models are insufficient or need improvement. 

\newpage
\section{Appendix A}

\hspace*{-1.5in}
\begin{figure}[H]
    \centerline{\includegraphics[width=16cm]{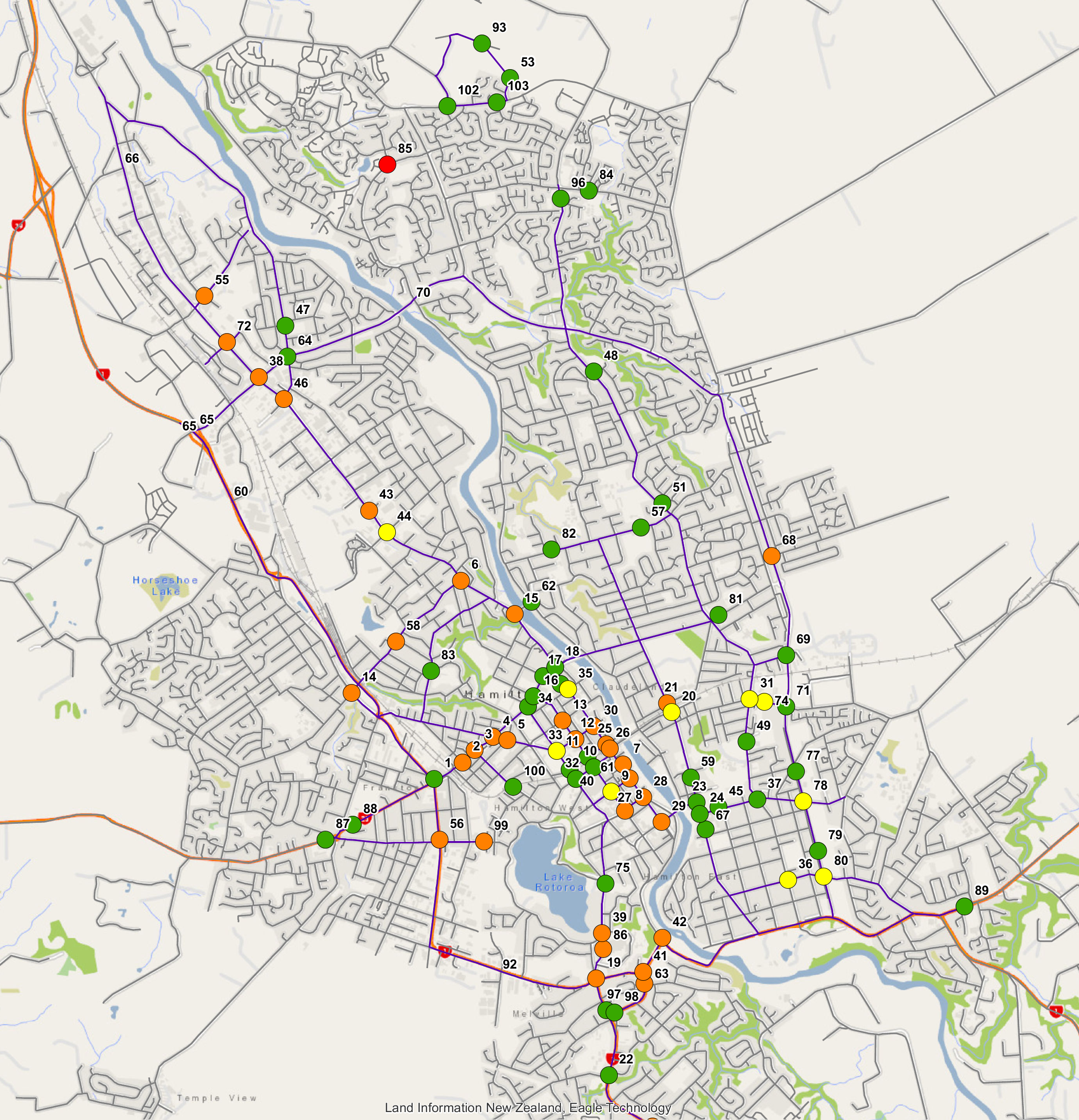}}
    \caption{Site Proximity to schools. Green is less than 300m, red is further than 800m}
\end{figure} 

\newpage
\section{Appendix B}

\hspace*{-1.5in}
\begin{figure}[H]
    \centerline{\includegraphics[width=14cm]{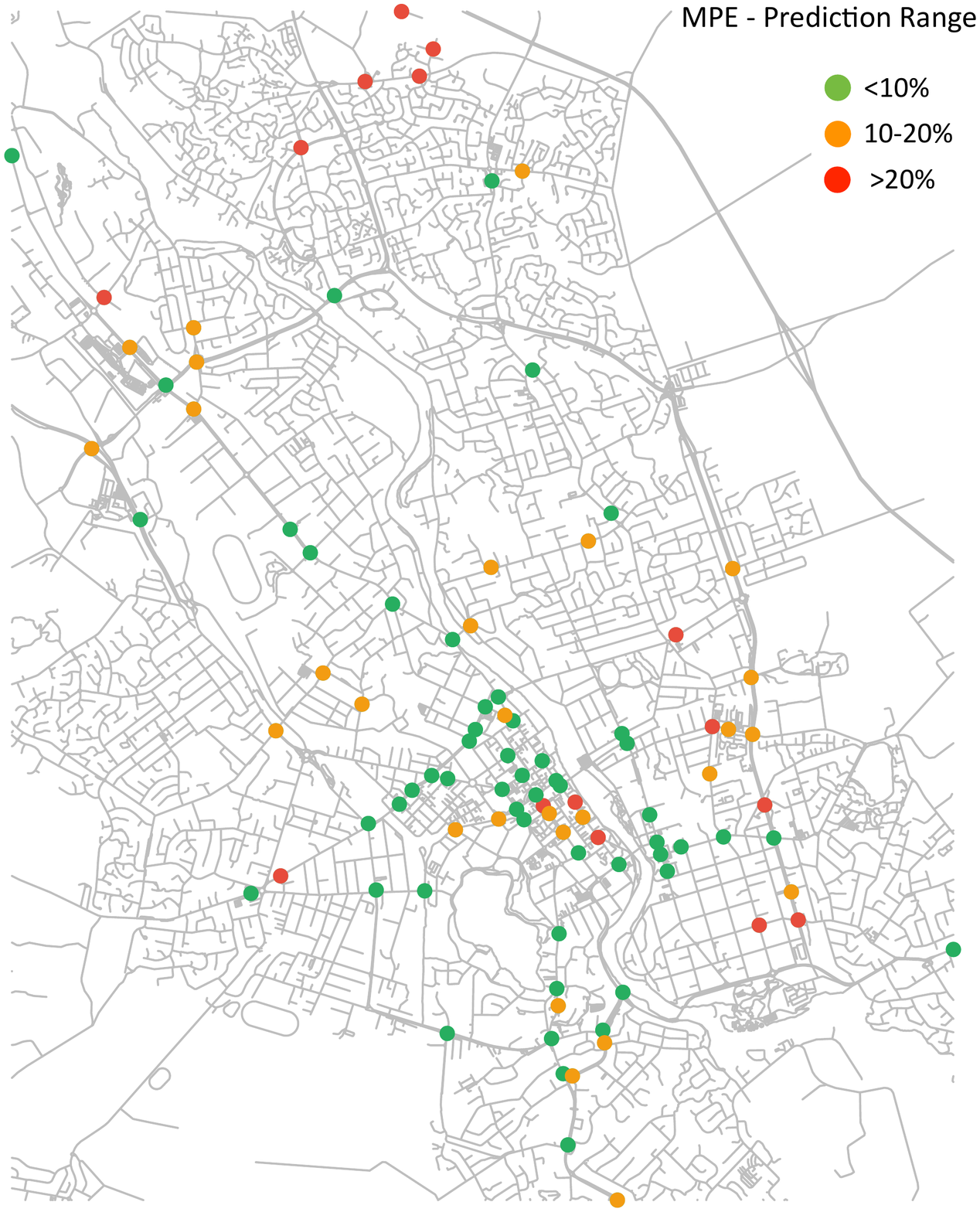}}
    \caption{MPE per Site for the prediction range (Model 11)}
\end{figure}

\end{document}